\documentclass{PoS}
\usepackage{epsfig}
\usepackage{subfigure}
\def \be {\begin{equation}}
\def \ee {\end{equation}}
\def \ba {\begin{eqnarray}}
\def \ea {\end{eqnarray}}
\title{Study of Hadron Scattering Using an Asymmetric Box}

\ShortTitle{Study of Hadron Scattering Using an Asymmetric Box}

\author{\speaker{C.~Liu}\thanks{For the CLQCD Collaboration}, Y. Shen, X.~Li, G.Z. Meng, X.~Feng, M.~Gong, S.~He\\
        School of Physics, Peking University, Beijing 100871, China\\
        E-mail: \email{liuchuan@pku.edu.cn, shendandan@gmail.com, blake@pku.edu.cn, gzhmeng@pku.edu.cn, pkuFengXu@pku.edu.cn, gongmingpku@gmail.com, fallenleave910@163.com}}

\author{Y. Chen, G. Li\\
        Institute of High Energy Physics, Academia Sinica, P.~O.~Box 918,  Beijing 100039, China\\
        E-mail: \email{cheny@mail.ihep.ac.cn, gli@mail.ihep.ac.cn}}

\author{Y.B. Liu, X.F. Meng\\
       Department of Physics, Nankai University, Tianjin 300071, China\\
        E-mail: \email{liuyb@nankai.edu.cn, mengxf@mail.nankai.edu.cn}}

\author{J.P. Ma\\
        Institute of Theoretical Physics, Academia Sinica, Beijing 100080, China\\
        E-mail: \email{majp@itp.ac.cn}}

\author{J.B. Zhang\\
       Department of Physics, Zhejiang University, Hangzhou 310027, China\\
       E-mail: \email{jbzhang@zimp.zju.edu.cn}}

\abstract{%
 We propose to study hadron-hadron scattering using
 lattice QCD in an asymmetric
 box which allows one to access more non-degenerate low-momentum modes for a given
 volume. The conventional L\"{u}scher's formula applicable in a symmetric box is
 modified accordingly. To illustrate the feasibility of this
 approach, pion-pion elastic scattering phase shifts in the $I=2$, $J=0$ channel
 are calculated within quenched approximation using improved gauge
 and Wilson fermion actions on anisotropic lattices in an asymmetric
 box. After the chiral and continuum extrapolation, we find that our
 quenched results for the scattering phase shifts
 in this channel are consistent with the experimental data
 when the three-momentum of the pion is below $300$MeV. Agreement is
 also found when compared with previous theoretical results from lattice
 and other means. Moreover, with the usage of asymmetric volume, we are able
 to compute the scattering
 phases in the low-momentum range (pion three momentum less than
 about $350$MeV in the center of mass frame)
 for over a dozen values of the pion three-momenta, much more than using
 the conventional symmetric box with comparable volume.}

\FullConference{The XXV International Symposium on Lattice Field Theory\\
         July 30-4 August 2007\\
         Regensburg, Germany}

\begin{document}

\section{Introduction}
 Hadron-hadron scattering experiments play a crucial role in understanding
 interaction among hadrons and QCD. Due to its {\em non-perturbative} feature,
 lattice QCD (LQCD) is the only systematic, non-perturbative
 method of QCD which in principle can be applied to calculate
 low-energy hadron-hadron scattering from first principles.
 This type of lattice calculations is realized by a finite size method proposed by
 L\"{u}scher~\cite{luscher91:finitea}.
 Using this technique, the scattering length and the scattering phase shifts
 for pion-pion scattering in the $I=2$, $J=0$ channel have been
 studied~\cite{gupta93:scat,fukugita95:scat,jlqcd99:scat,%
 JLQCD02:pipi_length,chuan02:pipiI2,juge04:pipi_length,chuan04:pipi,ishizuka05:pipi_length,%
 CPPACS03:pipi_phase,CPPACS04:pipi_phase_unquench,savage06:pipi}
 in both quenched and unquenched lattice QCD.

 However, non-degenerate low-momentum modes accessible
 for lattice simulation are usually limited due to computational cost.
 With cubic boxes, more non-degenerate low-momentum modes
 requires larger physical volumes. To circumvent this problem,
 we propose to use asymmetric boxes to study hadron-hadron scattering.
 The idea is tested in a quenched study of pion-pion scattering in the
 $I=2$, $J=0$ channel~\cite{chuan07:pipi} using improved anisotropic
 lattices.

 \section{L\"uscher's formulae extended to an asymmetric box}

 Consider a cubic box with size $L \times L \times L$ and periodic
 boundary condition. Three momentum
 of a single pion is quantized as: $\textbf{k}=(2\pi/L)\textbf{n}$
 with $\textbf{n}=(n_1,n_2,n_3)\in \mathbb{Z}^3$, where $\mathbb{Z}$ represents the
 set of all integers. For the two-pion system
 taking the center of mass reference frame, we define
 $\bar{\textbf{k}}^2$ of the pion pair in a box as:
 \begin{equation}
 \label{eq:kbar_def}
 E_{\pi\pi}=2\sqrt{m_{\pi}^2+\bar{\textbf{k}}^2}
 \end{equation}
 where $E_{\pi\pi}$ is the {\em exact} energy of a two-pion system
 with the two pions having three momentum $\textbf{k}$ and $-\textbf{k}$
 respectively in the center of mass frame.
 We further define $q^2$ via:
 \begin{equation}
 \label{eq:q_def}
 \bar{\textbf{k}}^2=\frac{4\pi^2}{L^2}q^2
 \end{equation}
 With these definitions, L\"uscher's formula for the $s$-wave scattering
 phase can be represented as:
  \begin{equation}
 \label{eq:original_luscher}
 \cot{\delta_0 (\bar{k})}=\frac{\mathcal{Z}_{00}(1,q^2)}{\pi^{3/2}q}
 \;,
 \end{equation}
 where the so-called zeta function $\mathcal{Z}_{lm}$ is given by:
 \begin{equation}
 \mathcal{Z}_{lm}(s,q^2)=\sum_{\textbf{n}}
 \frac{\mathcal{Y}_{lm}(\textbf{n})}{(\textbf{n}^2-q^2)^s}
 \;.
 \end{equation}
 In this definition,
 $\mathcal{Y}_{lm}(\textbf{r})=r^l\mathit{Y}_{lm}(\Omega_\textbf{r})$,
 with $Y_{lm}(\Omega_\textbf{r})$ being the usual spherical harmonics.

 We need to extend the above formulae to an asymmetric box of physical size:
 $L \times \eta_2 L \times \eta_3 L$.  The momentum in the box is quantized as:
 $\textbf{k}=(2\pi/L)\tilde{\textbf{n}}$ with
 $\tilde{\textbf{n}}\equiv(n_{1},n_{2}/\eta_2,n_{3}/\eta_3)$ and
 $\textbf{n}=(n_1,n_2,n_3)\in\mathbb{Z}^3$.
 Quantities $\bar{\textbf{k}}$ and
 $q^2$ are still defined according to Eq.~(\ref{eq:kbar_def}) and Eq.~(\ref{eq:q_def}).
 The formula for the scattering phase shifts is now modified
 to:~\cite{chuan04:asymmetric,chuan04:asymmetric_long}
 \begin{equation}
 \label{eq:luescher_modified}
 \cot{\delta_0
 (\bar{k})}= m_{00}(q)\equiv \frac{\mathcal{Z}_{00}(1,q^2;\eta_{2},\eta_{3})}{\pi^{3/2}\eta_{2}\eta_{3}q}
 \end{equation}
 with the {\em modified} zeta function $\mathcal{Z}_{lm}$ defined as:
 \begin{equation}
 \mathcal{Z}_{lm}(s,q^2;\eta_{2},\eta_{3})=
 \sum_{\textbf{n}}\frac{\mathcal{Y}_{lm}(\tilde{\textbf{n}})}{(\tilde{\textbf{n}}^2-q^2)^s}
 \end{equation}
 In the large volume limit, the formula for scattering length is:
 \be
 \label{eq:a0_asymmetric}
 E_{\pi\pi}-2m_{\pi}=-\frac{4\pi a_0}{\eta_2\eta_3m_\pi
 L^3}\left[1+c_1(\eta_2,\eta_3)\left(\frac{a_0}{L}\right)
 +c_2(\eta_2,\eta_3)\left(\frac{a_0}{L}\right)^2\right]+O(L^{-6})
 \;.
 \ee
 where the coefficients for the case $\eta_2=1$ and $\eta_3=2$, which is what we
 simulated in this work, are found to be:
 $c_1(1,2)=-1.805872$, $c_2(1,2)=1.664979$.

 \section{Simulation details and results}
 To test our idea of using the asymmetric box on hadron-hadron
 scattering, we perform a quenched study on the pion-pion
 scattering phase shift in the $I=2$, $J=0$ channel.
 In this section, we will briefly introduce our numerical results.
 More details can be found in Ref.~\cite{chuan07:pipi}.
 \begin{center}
 \begin{tabular}{|c||c|c|c|c|}
 \hline
 $\beta$  & Lattice & $a_s(GeV^{-1})$ & Confs.\\
 \hline
 2.080 & $8^2\times 16\times 40$ & 1.5677 & 464 \\
 \hline
 2.215 & $9^2\times 18\times  48$ & 1.3926 & 425 \\
 \hline
 2.492 & $12^2\times 24\times 64$ & 1.0459 & 105 \\
 \hline
 \end{tabular}
 \end{center}
  The gauge action used in this study is the tadpole improved gluonic
  action on anisotropic lattices~\cite{colin97,colin99} and
  the fermion action used in this calculation is the tadpole improved
  clover Wilson action on anisotropic
 lattice~\cite{klassen98:wilson_quark,klassen99:aniso_wilson,chuan01:tune,chuan06:tune_v,chuan06:tune_beta}.
 The anisotropy parameter $\xi=a_s/a_t$ is fixed to $5$ for all our lattices.
 Some of the simulation parameters are provided in the table above.

 To obtain the two-pion energy levels, we follow the standard
 procedure of correlation matrix measurements.
 For the single pion operators, we use:
 \be
 \pi^+(\textbf{x},t)=-\bar{d}(\textbf{x},t)r_5u(\textbf{x},t),
 \ee
 The Fourier transformed fields are defined accordingly.
 The $s$-wave two-pion operators in the $I=2$ channel are defined as:
 \begin{equation}
 \label{eq:twopion_def}
 \mathcal{O}_n(t)=\sum_{R}\pi^+_{R(\textbf{k}_n)}(t)\pi^+_{R(-\textbf{k}_n)}(t+1)
 \;.
 \end{equation}
 where the summation runs over all symmetry operations $R$ of the symmetry group
 for the asymmetric box. In the simulation, we measure the
 correlation matrix among different non-degenerate two-pion modes,
 using the two-pion operators defined in Eq.~(\ref{eq:twopion_def}):
 \begin{equation}
 \label{eq:correlation_matrix_def}
 \mathcal{C}_{mn}(t)=\langle\mathcal{O}^\dagger_m(t)\mathcal{O}_n(t_s)\rangle \;.
 \end{equation}
 We then constructed a new correlation
 matrix:~\cite{luscher90:finite}
 \begin{equation}
 \Omega(t,t_0)=\mathcal{C}(t_0)^{-\frac{1}{2}}\mathcal{C}(t)\mathcal{C}(t_0)^{-\frac{1}{2}}
 \end{equation}
  where $t_0$ is some suitable reference time.
  The eigenvalue $\lambda_i(t)$ of this new matrix $\Omega$ is:
 \begin{equation}
 \lambda_i(t,t_0)\propto e^{-E_i(t-t_0)}
 \end{equation}
 from which the two-pion energy eigenvalues $E_i$ are obtained.
  \begin{figure}[htb]
 \begin{center}
 \subfigure{\includegraphics[width=0.45\textwidth]{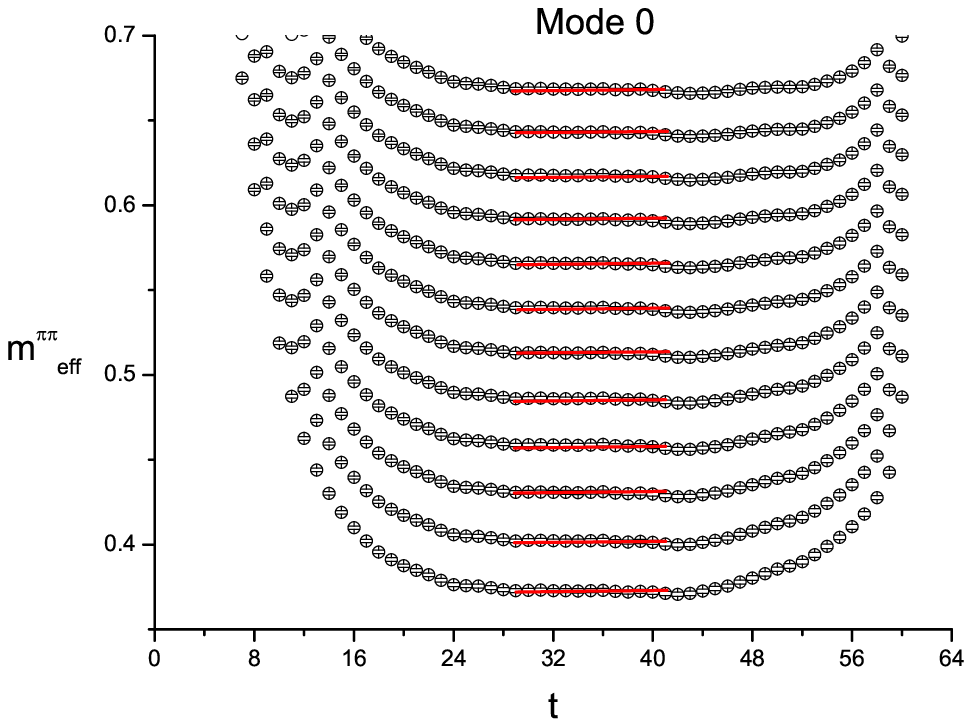}}
 \subfigure{\includegraphics[width=0.45\textwidth]{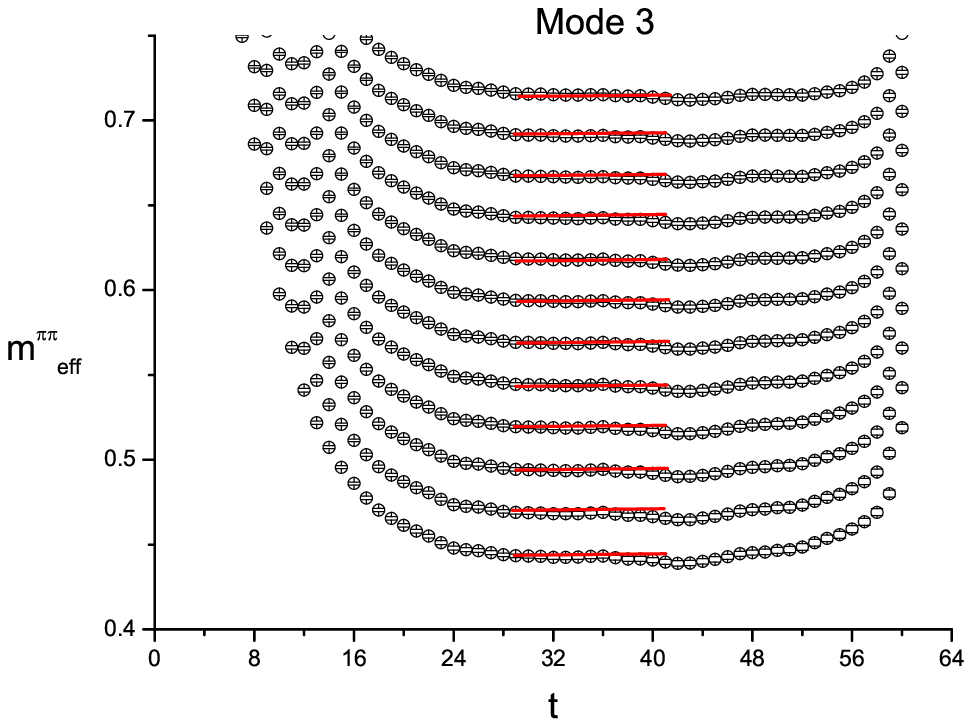}}
 \end{center}
 \caption{Effective mass plateaus of two-pion system at $\beta=2.492$
 after the diagonalization procedure. The horizontal line segments in
 the figure represent the fitting ranges of the plateaus. Results for
 other values of $\beta$ are similar. Only two momentum modes out of $6$ are
 shown.\label{fig:twopion_energy2492}}
 \end{figure}
 In Fig.~\ref{fig:twopion_energy2492}, we have shown the effective
 mass plateaus for the two-pion energies at $\beta=2.492$. The horizontal line segments
 designate the fitted values for the plateaus. Due to the usage of anisotropic
 lattices, good accuracy is achieved for two-pion energy levels.

 In our quenched study, the pion mass values are beyond the applicable range
 of chiral perturbation theory. We therefore adopted polynomial
 parametrization for the data points for the scattering lengths and the phase shifts.
 For the scattering length, we parameterize our data for $a_0/m_\pi$
 using either a linear function in $m^2_\pi$:
 \begin{equation}
 \frac{a_0}{m_{\pi}}=A_0+A_1m^2_{\pi};,
 \end{equation}
 or, for the case of $\beta=2.492$ where a curvature is seen, a quadratic form in $m^2_\pi$.
 \begin{figure}[htb]
 \begin{center}
 \subfigure{\includegraphics[width=0.48\textwidth]{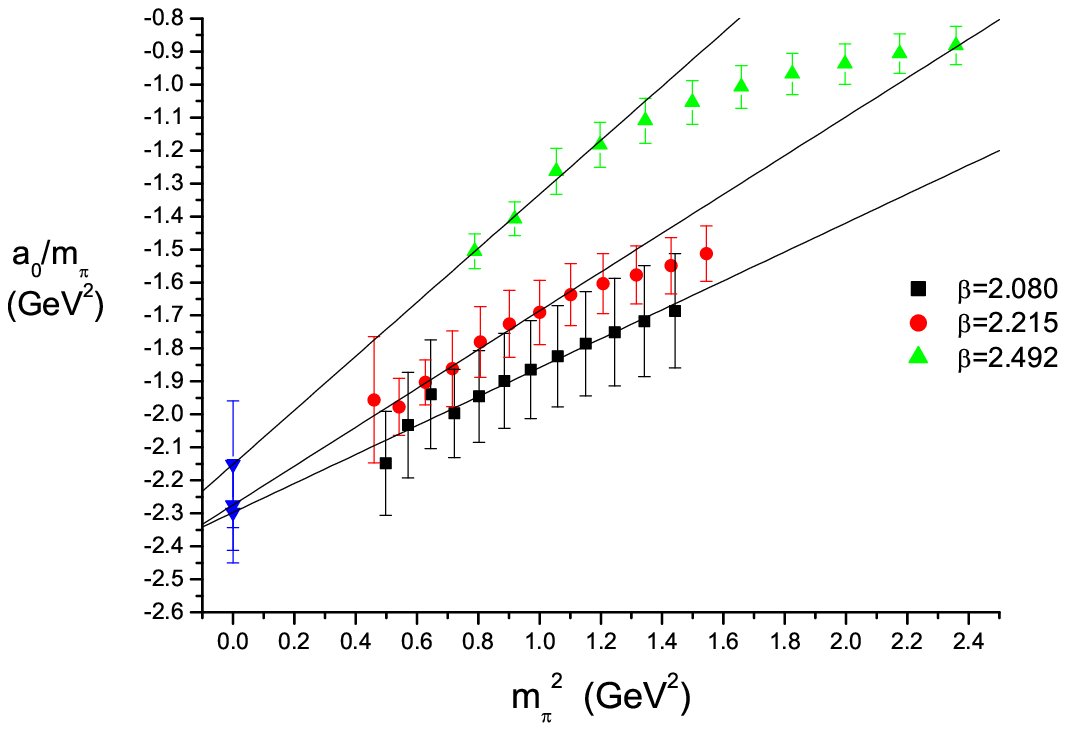}}
 \subfigure{\includegraphics[width=0.48\textwidth]{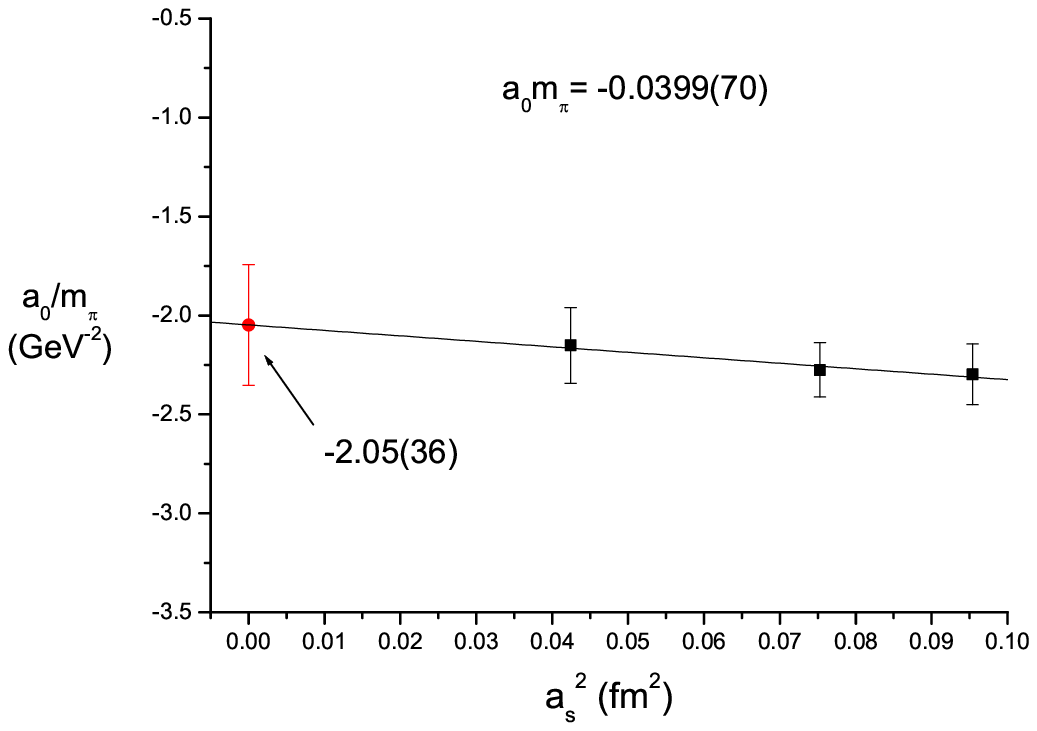}}
 \end{center}
 \caption{Left panel: chiral extrapolation of the quantity
 $\frac{a_0}{m_{\pi}}$.
 Right panel:  the corresponding continuum limit extrapolation of the
 quantity $\frac{a_0}{m_{\pi}}$. The extrapolated result is indicated
 by a solid circle near $a_s=0$.\label{fig:delta_a0_linear}}
 \end{figure}
 After the chiral extrapolation, the results for the scattering
 length at different lattice spacing values are extrapolated towards
 the continuum with a linear function in $a^2_s$. The chiral
 extrapolation and the continuum extrapolation for the scattering
 length are shown in Fig.~\ref{fig:delta_a0_linear}.
 As a result, we finally obtain: $a_0m_\pi=0.0359(59)$ which is compatible
 with results obtained in other lattice simulations, chiral
 perturbation theory and the experiment.

\begin{figure}[htb]
 \subfigure{\includegraphics[width=0.48\textwidth]{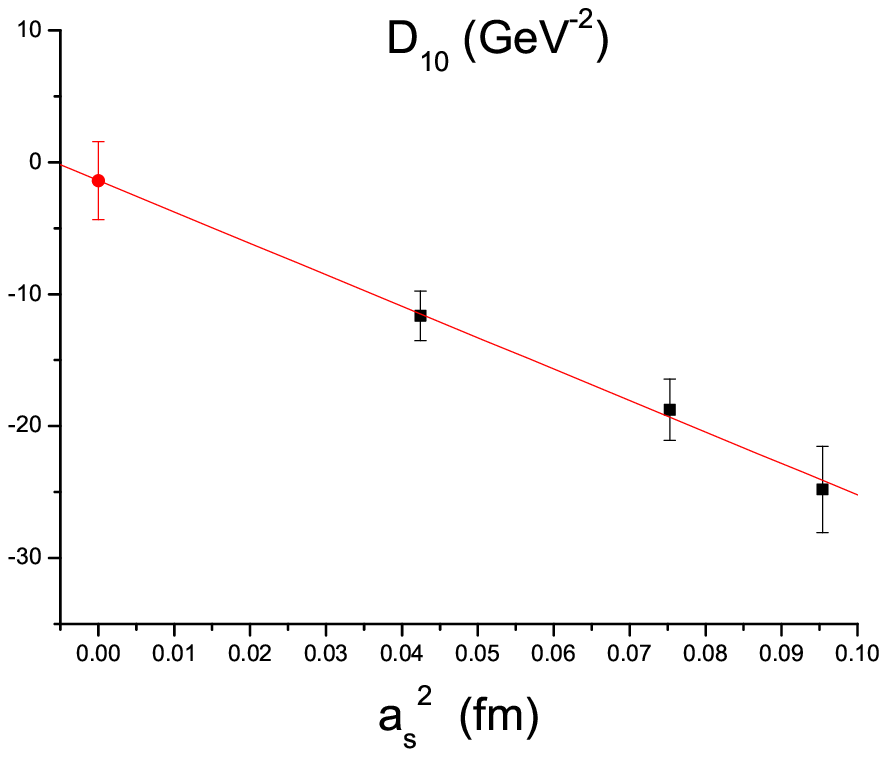}}
 \subfigure{\includegraphics[width=0.48\textwidth]{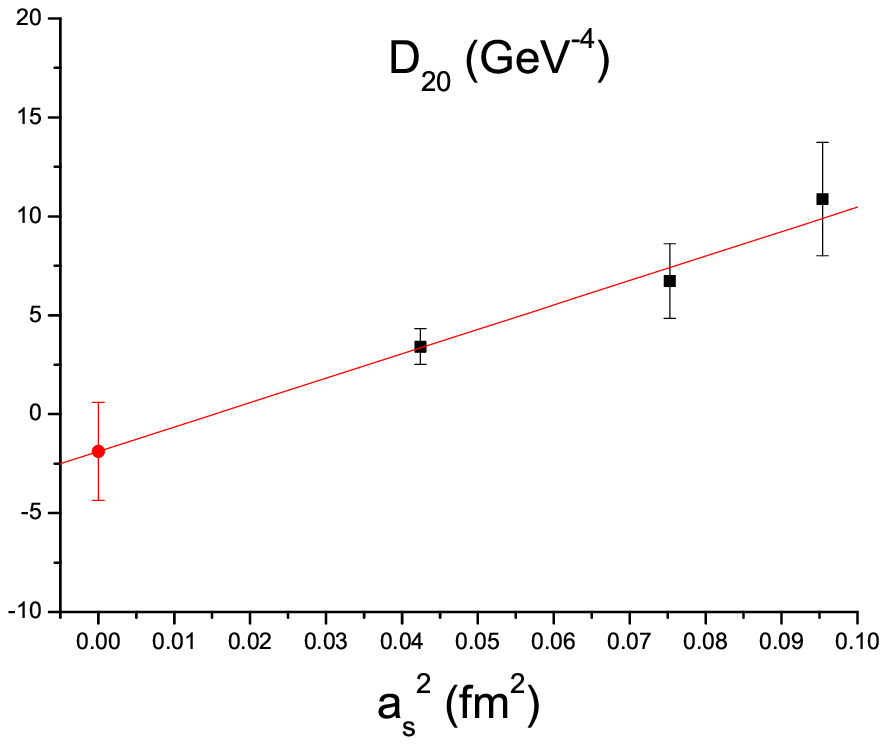}}
 \subfigure{\includegraphics[width=0.48\textwidth]{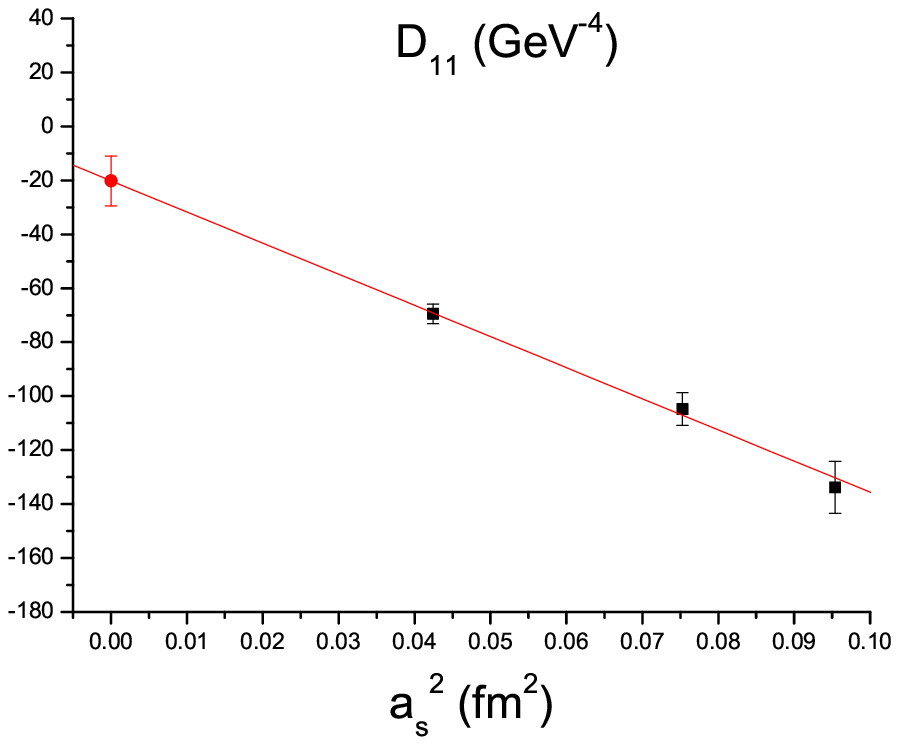}}
 \subfigure{\includegraphics[width=0.48\textwidth]{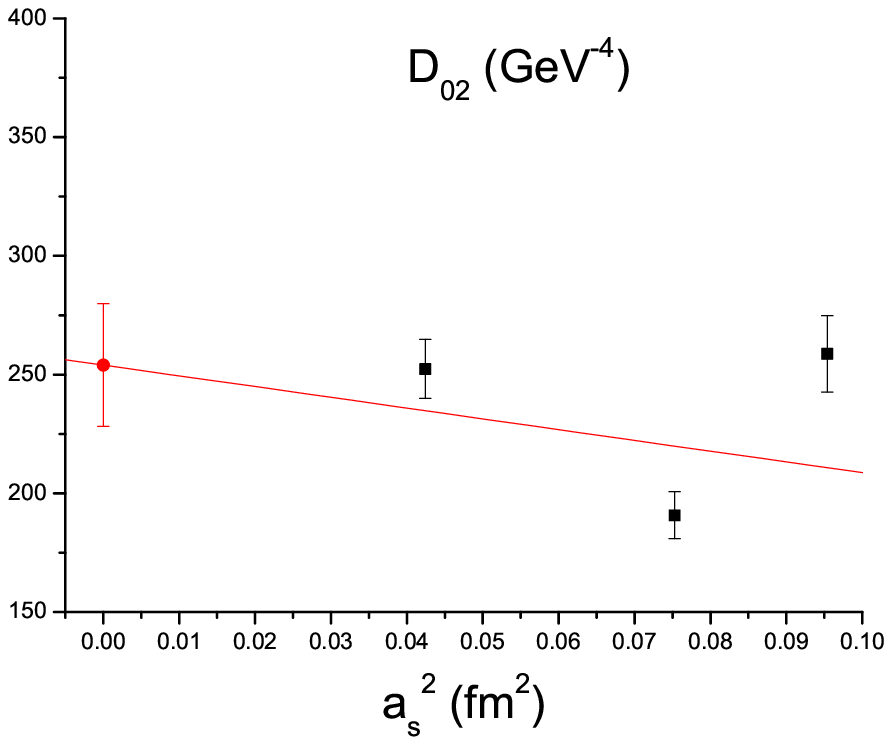}}
 \caption{The continuum limit extrapolation of the parameter
 $D_{ij}$. The solid circles near $a_s=0$ represent the corresponding
 continuum limit values.\label{fig:continuum_Dij}}
 \end{figure}

 Similarly,  we parameterize the phase shift
 $\delta$ itself by a polynomial in
 both $m^2_\pi$ and $\bar{k}^2$ as:
 \be
 \delta(m_{\pi}^2,\bar{k}^2) = D_{00}+D_{10}m_{\pi}^2
 +D_{20}m_{\pi}^4+D_{01}\bar{k}^2
 + D_{11}m_{\pi}^2\bar{k}^2+D_{02}\bar{k}^4
 \ee
 The coefficient $D_{00}$ vanishes due to chiral symmetry which is
 checked numerically as well. A linear extrapolation in $a^2_s$ is then
 performed for various coefficients $D_{ij}$ appearing in the above equation.
 Some of the extrapolations are shown in Fig.~\ref{fig:continuum_Dij}.

  \begin{figure}[htb]
 \begin{center}
 \subfigure{\includegraphics[width=0.7\textwidth]{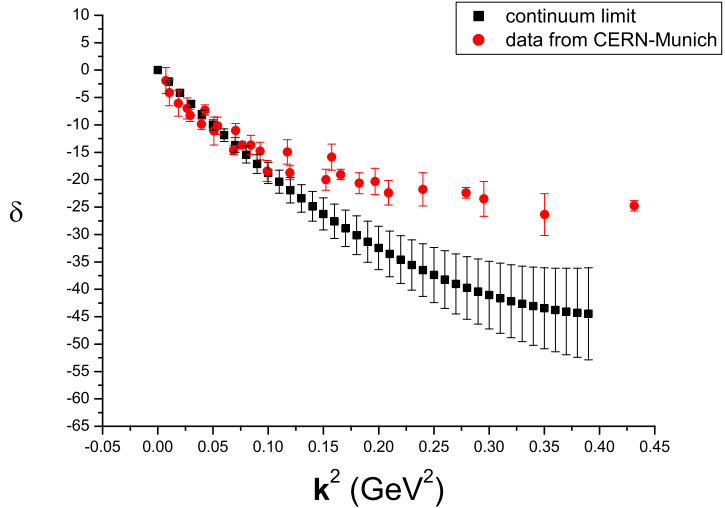}}
 \end{center}
 \caption{Comparison of our lattice results for the scattering phase
 shifts with the experimental data from
 CERN-Munich~\cite{Hoogland:1974cv}. Results are consistent with
 each other for $\textbf{k}^2$ below 0.1 GeV$^2$ which roughly
 corresponds to the center of mass energy of about
 $0.6$GeV.\label{fig:compare_experiment}}
 \end{figure}
 The results for the phase shift after the chiral and continuum
 extrapolations are summarized in Fig.~\ref{fig:compare_experiment}, together with the
 experimental results from CERN-Munich group~\cite{Hoogland:1974cv}.
 First, it is readily seen that, due to the usage of the asymmetric
 box, much more low-momentum modes are accessible with a relatively
 small volume, allowing much better comparison with the experimental data
 in the low-momentum range. Second, it is also found that our final results
 agree with the experimental results within errors for $\bar{\textbf{k}}^2$
 below $0.1$ GeV$^2$ which is about $\sqrt{s}=0.6$ GeV.
 At higher energies, our results deviate from the experimental
 results. This deviation might be caused by the systematic
 uncertainties in our calculation, e.g. quenching and chiral
 extrapolations. Numerically speaking, it is largely due to
 poor determination of the coefficient $D_{02}$.
 However, we would like to point out that, it is clear from our
 quenched calculation that,
 the asymmetric volume technique advocated here would also be
 useful for unquenched studies once
 the unquenched configurations become available.

\section{Conclusions}

 In this paper, we propose to study hadron-hadron scattering processes on
 lattices with asymmetric volume. This setup has the advantage that it provides
 much more non-degenerate low-momentum modes with a relatively small volume, allowing
 more detailed comparison both with the experiments and with other theoretical results.
 The feasibility of this idea is illustrated in a quenched
 calculation of pion-pion scattering length and scattering phases
 in the $I=2$, $J=0$ channel using clover improved lattice actions on anisotropic lattices.
 Our quenched results indicate that the usage of asymmetric volumes
 indeed allow us to access much more low-momentum modes than in the
 case of cubic volume of similar size.
 For $\bar{k}^2$ in the range  of $0.02$GeV$^2$ to $0.12$GeV$^2$, we have over
 a dozen of data points for the phase shift, much more than that in
 the cubic case with similar volume. It is also noted that, in the low-momentum
 region, after the chiral and continuum extrapolations, our results for the
 scattering length and the scattering phase shifts are in
 good agreement with the experimental data and are consistent with
 results obtained using other theoretical means.
 Although our calculation is now performed in the quenched approximation,
 similar calculations are also possible in the unquenched case once the gauge
 field configurations become available.

 \section*{Acknowledgments}
 The authors would like to thank Prof.~H.~Q.~Zheng and
 Prof.~S.~L.~Zhu of Peking University for helpful
 discussions. The numerical calculations were performed on DeepComp 6800 supercomputer of the
 Supercomputing Center of Chinese Academy of Sciences, Dawning 4000A
 supercomputer of Shanghai Supercomputing Center, and NKstar2
 Supercomputer of Nankai University.

\end{document}